\documentclass[12pt,final,a4paper,twoside]{iopart}
\expandafter\let\csname equation*\endcsname\relax
\expandafter\let\csname endequation*\endcsname\relax
\usepackage{amsmath}
\usepackage{amssymb}
\usepackage{mathtools}
\usepackage{mathbbol}
\usepackage{tensor}
\usepackage{cancel}
\usepackage{nicefrac}
\usepackage[T1]{fontenc}
\usepackage[utf8]{inputenc}
\usepackage{lmodern}
\usepackage{microtype}
\usepackage{xurl}
\usepackage[hidelinks]{hyperref}
\providecommand{\rmi}{\mathrm{i}}
\providecommand{\Eins}{\mathbf{1}}
\providecommand{\Tr}{\operatorname{tr}}
\providecommand{\tr}{\operatorname{tr}}
\newcommand{\LCd}{\mathcal{L}}
\newcommand{\onehalf}{{\textstyle\frac{1}{2}}}
\newcommand{\quarter}{{\textstyle\frac{1}{4}}}
\newcommand{\onethird}{{\textstyle\frac{1}{3}}}
\newcommand{\ihalf}{{\textstyle\frac{\rmi}{2}}}
\newcommand{\iquarter}{{\textstyle\frac{\rmi}{4}}}
\newcommand{\pfrac}[2]{\frac{\partial{#1}}{\partial{#2}}}
\newcommand{\ppfrac}[3]{\frac{\partial^{2}{#1}}{\partial{#2}\partial{#3}}}
\newcommand{\bgamma}{\pmb{\gamma}}
\newcommand{\bsigma}{\,\pmb{\sigma}}

\newcommand{\spc}{\,\pmb{\omega}}
\newcommand{\bk}{\pmb{R}}\let\br\bk
\newcommand{\rmD}{\mathrm{D}}
\newcommand{\bEins}{\pmb{\Eins}}

\newcommand{\afffias}{Frankfurt Institute for Advanced Studies (FIAS), Ruth-Moufang-Strasse~1,\\ 60438 Frankfurt am Main, Germany}
\newcommand{\affgsi}{GSI Helmholtzzentrum f\"ur Schwerionenforschung GmbH, Planckstrasse~1,\\ \,\,64291 Darmstadt, Germany}
\newcommand{\affjwg}{Goethe-Universit\"at, Max-von-Laue-Strasse~1, 60438 Frankfurt am Main, Germany}
\begin{document}
\title{Pauli-type coupling of spinors and curved spacetime}
\author{J~Struckmeier$^{1,2,3}$, D~Vasak$^1$, A~Redelbach$^1$, and H St\"ocker$^{1,2,3}$}
\address{$^1$\afffias}\address{$^2$\affjwg}\address{$^3$\affgsi\\[\medskipamount] 18~September~2026}
\ead{struckmeier@fias.uni-frankfurt.de (corresponding author)}
\begin{abstract}
We investigate Pauli-type interactions generated by minimal coupling of a specified covariantly regular representative of the Dirac Lagrangian.
A derivative-quadratic term proportional to a length parameter $\ell$ is a total divergence in free flat spacetime and leaves the free Dirac equation unchanged.
After electromagnetic coupling, it is equivalent, up to a divergence, to a Pauli operator with coefficient $-q\ell/6$.
We explain how a specified comparison between measured and predicted anomalous magnetic moments constrains this coefficient, separating historical illustrative scales from a conditional statistical bound.
In metric-compatible Riemann--Cartan spacetime, the same construction generates torsion-dependent derivative couplings and a spin-curvature coupling.
In the torsion-free limit, the latter reduces to the scalar mass coefficient $m+\ell R/12$.
On prescribed pure-axial and pure-vector torsion backgrounds, the characteristic form admits superluminal wavefronts relative to the spacetime metric in its nonzero-torsion Lorentzian regime; sufficiently large spacelike effective torsion instead leads to degeneracy and an ultrahyperbolic characteristic form.
For the observed cosmological constant and illustrative microscopic values of $\ell$, the vacuum mass correction is negligible.
High-curvature applications and the characteristic structure of the full coupled theory require a specified gravitational action and further dynamical analysis.
\end{abstract}
\maketitle
\section{Introduction}
Because of its \emph{linear} dependence on the derivatives of the spinor fields, the conventional Dirac Lagrangian is degenerate
in the sense that it does not provide a correlation of field derivatives to the canonical conjugate momentum fields.
The corresponding covariant Legendre map is rendered regular if the Dirac Lagrangian is amended by a term proposed by Gasiorowicz~\cite{gasiorowicz66} that is \emph{quadratic} in the derivatives of the spinor fields.
For dimensional reasons this additional term is necessarily associated with a length parameter $\ell$ and renders the Lagrangian non-degenerate.
We show in Sect.~\ref{sec:gasi-term} that this merely means to add a surface term to the conventional Dirac Lagrangian,
which, consequently, reproduces the free Dirac equation as the $\ell$-terms cancel when setting up the Euler-Lagrange field equations.

However, after minimally coupling spinor and electromagnetic gauge fields in Sect.~\ref{sec:minimal}, the interacting Lagrangian
gives the standard Quantum Electrodynamics (QED) Dirac equation plus an additional interaction term proportional to the length parameter $\ell$.
In Sect.~\ref{sec:equi} the interacting Dirac-Gasiorowicz Lagrangian is then shown to be, up to surface terms, equivalent to the Pauli Lagrangian~\cite{pauli41}.
The selection of the covariantly regular representative provides a structural motivation for the Pauli interaction.
It is an additional choice beyond gauge invariance: free Lagrangians differing by a divergence need not remain equivalent after application of the minimal-coupling prescription.
Related Dirac theories in Riemann--Cartan spacetime have been studied, for example, in Refs.~\cite{Fabbri:2014wda,Fabbri:2018hhv}.
Section~\ref{sec:bound} matches the Pauli coefficient to the anomalous magnetic moment, discusses the input dependence of numerical constraints, and distinguishes them from historical central-value estimates.

In the second part of the paper, Sections~\ref{Sec:Spacetime}--\ref{Sec:Discussion}, we study the analogous coupling to spacetime geometry.
The action must be invariant under coordinate changes and under local spin transformations covering the proper orthochronous Lorentz group.
The Lorentz part acts through the spin representation, while the spin connection provides the corresponding covariant derivative.
We use a first-order description in which tetrads and connection are initially independent, as in the covariant canonical framework of Ref.~\cite{vasak23}.
This choice does not uniquely determine the gravitational action, nor does the symmetry requirement alone specify couplings to arbitrary nonmetricity.
The explicit physical applications below are restricted to a compatible spin connection in metric-compatible Riemann--Cartan geometry.
Internal symmetries can also be included; canonical constructions for SU$(N)$ and Yang--Mills fields are discussed in Refs.~\cite{struckmeier08,struckmeier13}.

The length parameter in the selected Dirac representative gives rise to additional matter--gravity interactions, in analogy with the electromagnetic Pauli term.
The general terms consist of torsion-dependent derivative couplings and a spin-curvature coupling; they are not all scalar mass corrections.
For metric-compatible torsion-free spacetime, the spin-curvature coupling reduces to a scalar contribution proportional to the Ricci scalar.
A common coupling across different species is an additional assumption, not a consequence of the single-species calculation.

We also analyze the characteristic covectors of the generalized spinor equation on prescribed backgrounds, following the method of Velo and Zwanziger~\cite{PhysRev.186.1337,PhysRev.188.2218}.
This local cone analysis is distinguished from the characteristics and well-posedness of a complete dynamical gravitational--fermionic system.
\section{Non-degenerate Lagrangian of the Dirac field\label{sec:gasi-term}}
A non-degenerate Lagrangian for the Dirac field was proposed by Gasiorowicz~\cite{gasiorowicz66} to show that the indeterminacy
of a Lagrangian leads to ambiguities when formally applying the
``minimum coupling'' recipe\footnote{This should not be confused with the Gordon decomposition of the Dirac current to generate
the standard Pauli term $\propto q/m$, see, e.g., Hehl et al.~\cite{hehl91}.}.
It utilizes the fact that the Euler-Lagrange equation is unaffected by adding a divergence (surface) term to the Lagrangian.
The specific choice
\begin{equation}\label{eq:ld-dirac-regular}
\LCd_{0}=\frac{\rmi}{2}\left(\bar{\psi}\,\bgamma^{\alpha}\pfrac{\psi}{x^{\alpha}}-
\pfrac{\bar{\psi}}{x^{\alpha}}\bgamma^{\alpha}\psi\right)-m\,\bar{\psi}\psi+\pfrac{F^{\mu}}{x^{\mu}}\,,
\end{equation}
with
\begin{equation}\label{eq:gasi-term}
F^{\mu}=\frac{\rmi\ell}{6}\left(\bar{\psi}\bsigma^{\mu\beta}\pfrac{\psi}{x^{\beta}}+
\pfrac{\bar{\psi}}{x^{\beta}}\bsigma^{\beta\mu}\psi\right)\quad\Rightarrow\quad
\pfrac{F^{\mu}}{x^{\mu}}=\frac{\rmi\ell}{3}\pfrac{\bar{\psi}}{x^{\alpha}}\bsigma^{\alpha\beta}\pfrac{\psi}{x^{\beta}}\,,
\end{equation}
is associated with a length parameter $\ell$ for dimensional reasons.
$\LCd_{0}$ can be written in symmetric form as
\begin{align}
\LCd_{0}&=\frac{\rmi\ell}{3}\left(\pfrac{\bar{\psi}}{x^{\alpha}}-
\frac{\rmi}{2\ell}\bar{\psi}\,\bgamma_{\alpha}\right)\bsigma^{\alpha\beta}
\left(\pfrac{\psi}{x^{\beta}}+\frac{\rmi}{2\ell}\bgamma_{\beta}\psi\right)-\left(m-\ell^{-1}\right)\bar{\psi}\psi\,.
\label{eq:ld-dirac-coupling0}
\end{align}
This representation of the free Dirac Lagrangian suggests that the spinors are ``minimally coupled'' to the Dirac matrices, with the coupling constant $1/2\ell$.
The quadratic dependence of the Lagrangian~(\ref{eq:ld-dirac-coupling0}) on the derivatives of the spinors entails their unique correlation to the respective
conjugate polymomenta and thus renders the covariant Legendre map invertible for $\ell\ne0$---hence the classification ``non-degenerate.''
In four dimensions, the relevant matrix inverse can be checked explicitly:
\begin{equation*}
\bsigma^{\mu\nu}\left(\rmi g_{\nu\rho}\bEins
-\frac{\rmi}{3}\bgamma_\nu\bgamma_\rho\right)
=\delta^\mu_\rho\bEins.
\end{equation*}
This statement concerns all spacetime derivatives in the covariant Legendre map, not the ordinary equal-time velocity Hessian: $\bsigma^{00}=0$.
The latter retains its constrained character.

We use metric signature $(+,-,-,-)$ and natural units $\hbar=c=1$, restoring units in numerical estimates.
Our geometric conventions are specified below and in~\ref{app0-gd}; for general background see Ref.~\cite{misner73}.
The parameter $\ell$ is taken to be real and constant throughout the derivations.
The matrices $\bsigma^{\alpha\beta}$ are proportional to the Lorentz generators in the Dirac spinor representation and are defined by explicit gamma-matrix products:
\begin{equation}\label{eq:metric-def}
\bsigma^{\alpha\beta}=\frac{\rmi}{2}\left(\bgamma^{\alpha}\bgamma^{\beta}-\bgamma^{\beta}\bgamma^{\alpha}\right)\,,\quad
g^{\alpha\beta}\bEins=\frac{1}{2}\left(\bgamma^{\alpha}\bgamma^{\beta}+\bgamma^{\beta}\bgamma^{\alpha}\right)\,.
\end{equation}
For now $g^{\alpha\beta}$ is the (static) Minkowski metric, and $\bEins$ the unit matrix in spinor space.
Matrices in spinor space are typeset in boldface throughout this paper, with all spinor indices suppressed.

The Euler-Lagrange equations for the Lagrangian~(\ref{eq:ld-dirac-regular}) resp.~(\ref{eq:ld-dirac-coupling0}),
\begin{equation}\label{eq:elgl}
\pfrac{}{x^{\beta}}\pfrac{\LCd}{(\partial_{\beta}\psi)}-\pfrac{\LCd}{\psi}=0\,,\qquad
\pfrac{}{x^{\alpha}}\pfrac{\LCd}{(\partial_{\alpha}\bar{\psi})}-\pfrac{\LCd}{\bar{\psi}}=0\,,
\end{equation}
yield the standard Dirac equations in inertial spacetime as all terms depending on the length parameter $\ell$ cancel:
\begin{equation}\label{eq:dirac-equation}
\pfrac{\bar{\psi}}{x^{\alpha}}\rmi\bgamma^{\alpha}+m\bar{\psi}=0\,,\qquad\rmi\bgamma^{\beta}\pfrac{\psi}{x^{\beta}}-m\psi=0\,.
\end{equation}
The parameter $\ell$ does not affect the free bulk Dirac equation.
Adding the divergence changes the canonical momenta and may change the local canonical energy--momentum expression by improvement terms~\cite{hehl91}.
Under the same boundary prescription, this does not by itself imply different physical conserved charges.
In this work we select the representative by its covariant Legendre regularity, and then investigate the interactions produced by coupling that representative.
\section{\texorpdfstring{U$(1)$}{U(1)} symmetry -- minimal coupling to the abelian gauge field\label{sec:minimal}}
As the result of standard U$(1)$ gauge theory, a spinor couples minimally to the electromagnetic field $A_{\alpha}$ with the coupling constant (``charge'') $q$ via
\begin{equation}\label{eq:mincoupling}
\pfrac{\psi}{x^{\beta}}\mapsto\pfrac{\psi}{x^{\beta}}-\rmi qA_{\beta}\psi\eqqcolon\nabla_{\beta}\psi\,,\quad
\pfrac{\bar{\psi}}{x^{\alpha}}\mapsto\pfrac{\bar{\psi}}{x^{\alpha}}+\rmi q\,\bar{\psi}A_{\alpha}\eqqcolon\bar{\nabla}_{\alpha}\bar{\psi}\,
\end{equation}
provided that the spinors and the electromagnetic ``gauge field'' $A_\mu$ transform as
\begin{equation*}
\psi\mapsto\psi e^{\rmi\Lambda(x)},\qquad\bar{\psi}\mapsto\bar{\psi}e^{-\rmi\Lambda(x)},\qquad A_\mu\mapsto A_\mu+\frac{1}{q}\pfrac{\Lambda}{x^\mu}.
\end{equation*}
The gauge-invariant field strength is
\begin{equation*}
F_{\alpha\beta}=\pfrac{A_\beta}{x^\alpha}-\pfrac{A_\alpha}{x^\beta}.
\end{equation*}
The Maxwell term~$-\quarter F_{\alpha\beta}F^{\alpha\beta}$ specifies the free gauge-field dynamics; gauge invariance alone does not select a unique gauge-field action.

The system is thus described by a QED extension with a covariantly regular spinor sector
\begin{align}
\LCd_{1}&=\frac{\rmi\ell}{3}\left(\bar{\nabla}_{\alpha}\bar{\psi}-
\frac{\rmi }{2\ell}\bar{\psi}\,\bgamma_{\alpha}\right)\bsigma^{\alpha\beta}\left(
\nabla_{\beta}\psi+\frac{\rmi}{2\ell}\bgamma_{\beta}\psi\right)\nonumber\\
&\quad-\left(m-\ell^{-1}\right)\bar{\psi}\psi-\frac{1}{4}F^{\alpha\beta}F_{\alpha\beta}\,,
\label{eq:ld-dirac-coupling}
\end{align}
which expands to
\begin{align}
\LCd_{1}&=-\frac{\rmi}{2}\left(\bar{\nabla}_{\alpha}\bar{\psi}\right)\bgamma^{\alpha}\psi
+\frac{\rmi}{2}\bar{\psi}\,\bgamma^{\beta}\left(\nabla_{\beta}\psi\right)
-m\bar{\psi}\psi-\frac{1}{4}F^{\alpha\beta}F_{\alpha\beta}\nonumber\\
&\quad
+\frac{\rmi\ell}{3}\left(\bar{\nabla}_{\alpha}\bar{\psi}\right)
\bsigma^{\alpha\beta}\left(\nabla_{\beta}\psi\right).
\label{eq:ld-dirac-coupling2}
\end{align}
The last term proportional to $\ell$ emerges from the ``Gasiorowicz term''~\cite{gasiorowicz66} in the Lagrangian~(\ref{eq:ld-dirac-regular}).
In contrast to the Lagrangian interaction term proposed by Pauli~\cite{pauli41}, the ``Gasiorowicz term'' now entails
a \emph{minimally} coupled anomalous interaction of the spin with the electromagnetic field $A_\alpha$ in~(\ref{eq:ld-dirac-coupling2}).

\begin{subequations}\label{eq:el-dirac-gi}
\begin{align}
\rmi\bgamma^{\alpha}\,\nabla_{\alpha}\psi-m\,\psi-\frac{q\ell}{6}\bsigma^{\alpha\beta}F_{\alpha\beta}\,\psi&=0\\
\bar{\nabla}_{\alpha}\bar{\psi}\,\rmi\bgamma^{\alpha}+m\,\bar{\psi}+\frac{q\ell}{6}\bar{\psi}\bsigma^{\alpha\beta}F_{\alpha\beta}&=0\,.
\end{align}
\end{subequations}
We conclude that the ``Gasiorowicz term'', i.e., the last term of Eq.~(\ref{eq:ld-dirac-regular})---which does not modify
the field equation of the non-interacting fermion system---actually modifies the field equation of the minimally coupled system~(\ref{eq:ld-dirac-coupling2}).
The coupling constant $\ell$ that dropped out for the free Dirac fields, now becomes a physical, emergent length parameter.

The local Pauli operator has mass dimension five, and its coefficient therefore has negative mass dimension.
It is nonrenormalizable by perturbative power counting~\cite{weinberg96}, but can be treated as an interaction in an effective field theory.
A non-Gaussian ultraviolet fixed point is a different possibility: functional renormalization-group studies provide evidence for such fixed points in truncations containing a Pauli interaction~\cite{Gies:2020xuh,GiesTam2024}.
These results motivate further investigation, but do not establish that a nonzero fundamental Pauli coefficient is a necessary ingredient of QED or that the present local theory has a demonstrated ultraviolet completion.
\section{Pauli Lagrangian via minimal coupling\label{sec:equi}}
Integrating by parts gives the following identity modulo a total divergence, denoted by $\simeq$.
The corresponding actions have the same bulk Euler--Lagrange equations for compactly supported variations or appropriate boundary conditions:
\begin{equation*}
\frac{\rmi\ell}{3}\left(\bar{\nabla}_{\alpha}\bar{\psi}\right)
\bsigma^{\alpha\beta}\left(\nabla_{\beta}\psi\right)
\simeq-\frac{\rmi\ell}{3}\,\bar{\psi}\bsigma^{\alpha\beta}\,\nabla_{\alpha}\nabla_{\beta}\,\psi
=-\frac{q\ell}{6}\,F_{\alpha\beta}\,\bar{\psi}\,\bsigma^{\alpha\beta}\psi \,.
\end{equation*}
Here the antisymmetry of $\bsigma^{\alpha\beta}$ and
\begin{equation*}
(\nabla_\alpha\nabla_\beta-\nabla_\beta\nabla_\alpha)\psi
=-\rmi qF_{\alpha\beta}\psi
\end{equation*}
have been used.
The resulting operator has the form of the Pauli interaction~\cite{pauli41}.
It is nonminimal when classified as an operator added to the conventional Dirac--Maxwell Lagrangian, but here arises from minimal coupling of the selected derivative-quadratic representative.
The interaction terms by Gasiorowicz, $\LCd_{\mathrm{int}}$, and Pauli, $\LCd_{\mathrm{int, Pauli}}$, thus differ only by a surface term
\begin{equation*}
\LCd_{\mathrm{int}}-\LCd_{\mathrm{int, Pauli}}=\pfrac{F^\alpha}{x^\alpha},
\end{equation*}
which is a generalization of Eq.~(\ref{eq:gasi-term}) for $A_\beta\not\equiv0$:
\begin{equation*}
F^\mu=\frac{\rmi\ell}{6}\left[\bar{\psi}\bsigma^{\mu\beta}\left(\pfrac{\psi}{x^{\beta}}
-\rmi q\,A_{\beta}\psi\right)+\left(\pfrac{\bar{\psi}}{x^{\beta}}
+\rmi q\,\bar{\psi}A_{\beta}\right)\bsigma^{\beta\mu}\psi\right].
\end{equation*}
The two interaction densities therefore give the same bulk equations when their boundary terms are treated consistently.
For completeness, variation with respect to the electromagnetic potential in this equivalent representation yields
\begin{equation*}
\partial_\mu\left(F^{\mu\nu}+\frac{q\ell}{3}\bar\psi\bsigma^{\mu\nu}\psi\right)
=-q\bar\psi\bgamma^\nu\psi.
\end{equation*}
This displays the dipole contribution to the electromagnetic source.
The sign on the right follows from the convention $\nabla_\mu=\partial_\mu-\rmi qA_\mu$.
Our subsequent gravitational analysis focuses on the spinor equation; a closed matter--gravity model requires a specified gravitational action.
\section{Emergent length and anomalous magnetic moment\label{sec:bound}}
The parameter $\ell$ fixes a tree-level contribution to the magnetic dipole operator.
It must be distinguished from the loop-induced Pauli form factor already present in ordinary QED, whose leading contribution was calculated by Schwinger~\cite{schwinger48}.
Writing the physical magnetic moment operator as
\begin{equation}
\boldsymbol\mu=g\frac{q_{\rm phys}}{2m}\boldsymbol s,
\qquad a=\frac{g-2}{2},
\label{eq:magnetic-moment-definition}
\end{equation}
with $\boldsymbol s$ the spin operator, distinguishes the coefficient multiplying spin from the moment component of a polarized spin-$1/2$ state.
With the convention $F_{\mu\nu}=\partial_\mu A_\nu-\partial_\nu A_\mu$ and the source equation above, $q_{\rm phys}=-q$.
The magnitude of the anomalous Pauli-operator coefficient is $|q_{\rm phys}|\,|a_\ell|/(4m)$.
Matching it to the coefficient $|q\ell|/6$ gives
\begin{equation}
\frac{|q\ell|}{6}=\frac{|q|\,|a_\ell|}{4m},
\qquad |a_\ell|=\frac{2m|\ell|}{3},
\qquad |\ell|=\frac{3\hbar}{2mc}|a_\ell|.
\label{eq:pauli-moment-matching}
\end{equation}
The last expression restores units. It is an operator matching, not an identification of the Lagrangian coefficient with the entire magnetic moment.
A signed fit must use a consistent charge and Pauli-coefficient convention; the magnitude relations suffice for the size estimates below.

As historical benchmarks, retain the discrepancy magnitudes used in the original 2024 comparison:
$|\Delta a_\mu|=2.49\times10^{-9}$ and $|\Delta a_e|=8.76\times10^{-13}$.
The muon benchmark reflects the theory and experimental inputs discussed in Refs.~\cite{Logashenko:2018pcl,Aoyama:2020ynm,Hoecker:CERN2023,Muong-2:2023cdq}.
The earlier electron comparison is associated with Refs.~\cite{Electron_atoms7010028,Electron_Harvard_meas} and depends on the adopted independent determination of the fine-structure constant.
Using these central-value magnitudes only as illustrative contributions, together with the CODATA 2022 masses and $\hbar c$~\cite{CODATA2022}, gives
\begin{equation}
|\ell_\mu|_{\rm hist}\simeq6.98\times10^{-9}\,\mathrm{fm},
\qquad
|\ell_e|_{\rm hist}\simeq5.07\times10^{-10}\,\mathrm{fm}.
\label{eq:historical-length-scales}
\end{equation}
These are not confidence-level upper bounds.

To specify a statistical constraint, let $d=a_{\rm exp}-a_{\rm SM}$ and let $\sigma$ denote its combined standard uncertainty.
In a single-parameter Gaussian treatment, absent cancellation by other new contributions, $|a_\ell-d|\leq k\sigma$ implies the conservative magnitude bound
\begin{equation*}
|\ell|\leq\frac{3\hbar}{2mc}\bigl(|d|+k\sigma\bigr).
\end{equation*}
The theoretical inputs, correlations, and confidence convention must be identified in using this expression.

For an updated, explicitly specified example, the revised 2025 Muon $g-2$ Theory Initiative analysis adopts a consolidated lattice-QCD leading hadronic-vacuum-polarization input and quotes
$d_\mu=38(63)\times10^{-11}$~\cite{MuonWP2025}.
The disagreement among data-driven hadronic inputs is part of the motivation for that choice, so this residual is not interchangeable with the historical one.
Under the Gaussian assumptions above, $k=1.96$ gives the illustrative bound $|\ell_\mu|\lesssim4.5\times10^{-9}\,\mathrm{fm}$.
The detailed final Fermilab experimental report was submitted in June 2026~\cite{MuonFinal2026}.
These stated inputs define the example, rather than an exhaustive or model-independent present limit.
For electrons, a more precise magnetic-moment measurement was published in 2023~\cite{Electron2023}; a revised bound must specify which independent $\alpha$ determination enters $a_{\rm SM}$ and propagate its uncertainty.

Loop corrections generated by a fundamental Pauli coefficient require a specified quantum and renormalization prescription.
The numerical suppression obtained in the extended-object quantization of Ref.~\cite{Sastry:1999is} is not a universal estimate for the local theory considered here.
No model-independent $10^{-22}$ suppression of higher-order effects is assumed.
\section{Coupling to spacetime}\label{Sec:Spacetime}
In this section we examine the system of fermions in curved spacetime rather than the Lorentz invariant electrodynamics considered so far.
The curved geometry of spacetime is furnished by a generic metric $g_{\mu\nu}(x)$, with the same signature as the flat Minkowski spacetime,
that is expressed locally by orthonormal vierbein (tetrad) fields $e\indices{_i^\mu}(x)$.
The local calculations below do not require a single global tetrad.
Where a global spinor field is considered, a suitable spin structure is assumed.
The fundamental spin matrices $\bgamma^\mu(x)$ are built from the tetrads
for any representation of the static Dirac matrices $\bgamma^i$ according to $\bgamma^\mu(x)=\bgamma^i\,e\indices{_i^\mu}(x)$,
where $\mu=0,\ldots,3$ is the coordinate index in curved spacetime and $i=0,\ldots,3$ the Lorentz index in the locally flat tangent space.
Correspondingly, we have $\bsigma^{\alpha\beta}(x)=\bsigma^{kj}\,e\indices{_k^\alpha}(x)\,e\indices{_j^\beta}(x)$.
The transition of the Dirac equations~(\ref{eq:dirac-equation}) in a fixed flat spacetime background to the Dirac equations in a dynamic
curved spacetime can be summarized by applying the minimal coupling rules~\cite{frankel01,struckmeier21a}
\begin{equation}\label{eq:spincovderi}
\pfrac{\psi}{x^\beta}\rightarrow\pfrac{\psi}{x^\beta}+\spc_{\beta}\,\psi\eqqcolon\rmD_{\beta}\psi,\quad
\pfrac{\bar{\psi}}{x^\alpha}\rightarrow\pfrac{\bar{\psi}}{x^\alpha}-\bar{\psi}\,\spc_{\alpha}\eqqcolon\bar{\rmD}_{\alpha}\bar{\psi},
\end{equation}
with the \emph{spinor connection},
\begin{equation}\label{eq:spinorconnection}
\spc_{\alpha}(x)=-\frac{\rmi}{4}\bsigma^{kj}\omega_{kj\alpha}(x),
\end{equation}
where $\omega_{kj\alpha}(x)$ are the \emph{spin connection} coefficients~\cite{wheeler57,frankel01,yepez08}.
The partial derivatives of the spinors in the non-degenerate Lagrangian~(\ref{eq:ld-dirac-coupling0}) are substituted according
to the transformation rules~(\ref{eq:spincovderi}) by covariant derivatives.
Within the canonical construction of Ref.~\cite{struckmeier21a}, these rules implement coupling to the chosen spin connection.
Here they define the covariantization of the selected matter Lagrangian; they do not by themselves fix the gravitational action.
The spinor connection displayed in Eq.~(\ref{eq:spinorconnection}) contains the Lorentz-algebra part of the connection.
Before metric compatibility is imposed, the divergence rewritings below are formal identities with an affine connection; they are not a unique prescription for coupling spinors to arbitrary nonmetricity.
The explicit physical analysis from the next section onward uses the compatible spin connection and a metric-compatible Riemann--Cartan background.
With $g(x)$ denoting the determinant of the now \emph{dynamic} metric $g_{\alpha\beta}(x)$, this yields the gauge-covariant
(locally spin- and diffeomorphism-invariant) Lagrangian density $\tilde{\LCd}_{2}$ with a covariantly regular spinor sector
\begin{align*}
\tilde{\LCd}_{2}&=\left[\frac{\rmi\ell}{3}\!\left(\bar{\rmD}_{\alpha}\bar{\psi}
-\frac{\rmi }{2\ell}\bar{\psi}\,\bgamma_{\alpha}\right)\bsigma^{\alpha\beta}\!
\left(\!\rmD_{\beta}\psi+\frac{\rmi}{2\ell}\bgamma_{\beta}\psi\right)
-\left(m-\ell^{-1}\right)\!\bar{\psi}\psi\right]\,\sqrt{-g}+\tilde{\LCd}_{\text{Gr}}.
\end{align*}
As marked by the tilde, $\tilde{\LCd}_{2}$ represents a relative scalar of weight $w=1$ (i.e., a scalar density), rather than an absolute scalar.
The Lagrangian $\tilde{\LCd}_{\text{Gr}}\big(\partial\spc,\spc,\partial\bgamma,\bgamma\big)$ denotes the \emph{scalar density} Lagrangian
for the gravitational sector, expressed here schematically in terms of the spinor connection and spacetime-dependent $\bgamma^\mu$ matrices.
It is left unspecified; consequently no regularity claim for the complete gravitational Legendre map, or set of coupled gravitational field equations, is made.

$\tilde{\LCd}_{2}$ expands to
\begin{align}
\tilde{\LCd}_{2}&=\left[\frac{\rmi}{2}\bar{\psi}\,\bgamma^{\beta}\left(\rmD_{\beta}\psi\right)
-\frac{\rmi}{2}\left(\bar{\rmD}_{\alpha}\bar{\psi}\right)\bgamma^{\alpha}\psi
-m\,\bar{\psi}\psi+\frac{\rmi\ell}{3}\left(\bar{\rmD}_{\alpha}\bar{\psi}\right)
\bsigma^{\alpha\beta}\left(\rmD_{\beta}\psi\right)\right]\sqrt{-g}+\tilde{\LCd}_{\text{Gr}}.
\label{eq:ld-dirac-regular-gauged-expanded}
\end{align}
The Euler-Lagrange equation~(\ref{eq:elgl}) for the spinor $\psi$ follows as:
\begin{align}
&\left[\rmi\bgamma^{\beta}-\frac{\rmi\ell}{3}\left(\pfrac{\bsigma^{\alpha\beta}}{x^{\alpha}}+\spc_{\alpha}\bsigma^{\alpha\beta}-\bsigma^{\alpha\beta}\spc_{\alpha}
+\bsigma^{\alpha\beta}\frac{1}{\sqrt{-g}}\pfrac{\sqrt{-g}}{x^{\alpha}}\right)\right]\rmD_{\beta}\psi\nonumber\\
-&\left[m-\frac{\rmi}{2}\left(\pfrac{\bgamma^{\alpha}}{x^{\alpha}}+\spc_{\alpha}\bgamma^{\alpha}-\bgamma^{\alpha}\spc_{\alpha}
+\bgamma^{\alpha}\frac{1}{\sqrt{-g}}\pfrac{\sqrt{-g}}{x^{\alpha}}\right)
+\frac{\rmi\ell}{3}\bsigma^{\alpha\beta}\left(\pfrac{\spc_{\beta}}{x^{\alpha}}+\spc_{\alpha}\,\spc_{\beta}\right)\right]\psi=0.
\label{eq:gen-dirac-equation0}
\end{align}
It describes the spinor dynamics in curved spacetime with the spacetime-dependent $\bgamma^{\mu}$-matrices and spinor connection $\spc_{\mu}$.
The expressions proportional to $\ell$ emerge from the ``Gasiorowicz term'',
i.e., the last term of the Lagrangian~(\ref{eq:ld-dirac-regular-gauged-expanded}).

Remarkably, the sum proportional to the skew-symmetric $\bsigma^{\alpha\beta}$ in the last term of~(\ref{eq:gen-dirac-equation0})
yields a direct coupling of $\psi$ to the curvature spinor-tensor $\bk_{\alpha\beta}$,
\begin{equation}\label{eq:curvature-spinor}
\bk_{\alpha\beta}=\pfrac{\spc_{\beta}}{x^{\alpha}}-\pfrac{\spc_{\alpha}}{x^{\beta}}+\spc_{\alpha}\,\spc_{\beta}-\spc_{\beta}\,\spc_{\alpha}\,,
\end{equation}
which is in fact a $(1,1)$-spinor-$(0,2)$-tensor.
Moreover, the coupling term $\ell\bsigma^{\alpha\beta}\bk_{\alpha\beta}\,\psi$ coincides \emph{in its form} exactly with the Pauli coupling term
$\ell\bsigma^{\alpha\beta}F_{\alpha\beta}\,\psi$ from Eq.~(\ref{eq:el-dirac-gi})---yet here summation over the spinor indices of $\bk_{\alpha\beta}$ is understood in addition.

To discuss the terms in brackets of Eq.~(\ref{eq:gen-dirac-equation0}), we first note that $\sqrt{-g}$,
with $g$ the determinant of the covariant metric $g_{\mu\nu}$, is a relative scalar of weight $w=1$.
Our affine-connection convention places the differentiation index last:
\begin{equation*}
X^\beta{}_{;\alpha}=\partial_\alpha X^\beta+\Gamma\indices{^\beta_\xi_\alpha}X^\xi.
\end{equation*}
The covariant derivative of the volume density then follows as
\begin{equation*}
{\left(\sqrt{-g}\,\right)}_{;\alpha}=\pfrac{\sqrt{-g}}{x^\alpha}-\sqrt{-g}\,\Gamma\indices{^\beta_\beta_\alpha}.
\end{equation*}
With $S\indices{^{\,\beta}_{\alpha\xi}}=\Gamma\indices{^{\,\beta}_{[\alpha\xi]}}$ denoting the Cartan torsion tensor,
the sums in brackets can thus be expressed in terms of covariant divergences $\bgamma\indices{^\alpha_;_\alpha}$,
\begin{align*}
&\quad\;\pfrac{\bgamma^{\alpha}}{x^{\alpha}}+\spc_{\alpha}\bgamma^{\alpha}-\bgamma^{\alpha}\spc_{\alpha}
+\bgamma^{\alpha}\frac{1}{\sqrt{-g}}\pfrac{\sqrt{-g}}{x^{\alpha}}\\
&=\bgamma\indices{^\alpha_;_\alpha}+\frac{1}{\sqrt{-g}}{\left(\sqrt{-g}\right)}_{;\alpha}\bgamma^\alpha
-\bgamma^\alpha\left(\Gamma\indices{^\beta_\alpha_\beta}-\Gamma\indices{^\beta_\beta_\alpha}\right)\\
&=\frac{1}{\sqrt{-g}}\tilde{\bgamma}\indices{^\alpha_;_\alpha}-2\bgamma^\alpha\,S\indices{^\beta_\alpha_\beta}
\end{align*}
and $\bsigma\indices{^\alpha^\beta_;_\alpha}$
\begin{align*}
&\quad\;\pfrac{\bsigma^{\alpha\beta}}{x^{\alpha}}+\spc_{\alpha}\bsigma^{\alpha\beta}-\bsigma^{\alpha\beta}\spc_{\alpha}
+\bsigma^{\alpha\beta}\frac{1}{\sqrt{-g}}\pfrac{\sqrt{-g}}{x^{\alpha}}\\
&=\bsigma\indices{^\alpha^\beta_;_\alpha}+\frac{1}{\sqrt{-g}}{\left(\sqrt{-g}\right)}_{;\alpha}\bsigma^{\alpha\beta}
-\bsigma^{\xi\beta}\left(\Gamma\indices{^\alpha_\xi_\alpha}-\Gamma\indices{^\alpha_\alpha_\xi}\right)
-\bsigma^{\alpha\xi}\Gamma\indices{^\beta_\xi_\alpha}\\
&=\frac{1}{\sqrt{-g}}\tilde{\bsigma}\indices{^\alpha^\beta_;_\alpha}-2\bsigma^{\xi\beta}S\indices{^\alpha_\xi_\alpha}+\bsigma^{\xi\alpha}S\indices{^\beta_\xi_\alpha}.
\end{align*}
The Euler-Lagrange equation~(\ref{eq:gen-dirac-equation0}) is thus equivalently written in terms of the torsion tensor and the curvature spinor-tensor as the manifest tensor equation:
\begin{align}
&\left[\rmi\bgamma^{\beta}-\frac{\rmi\ell}{3}\left(\frac{1}{\sqrt{-g}}\tilde{\bsigma}\indices{^\alpha^\beta_;_\alpha}
-2\bsigma^{\xi\beta}S\indices{^\alpha_\xi_\alpha}+\bsigma^{\xi\alpha}S\indices{^\beta_\xi_\alpha}\right)\right]\rmD_{\beta}\psi\nonumber\\
-&\left[m-\frac{\rmi}{2}\left(\frac{1}{\sqrt{-g}}\tilde{\bgamma}\indices{^\alpha_;_\alpha}-2\bgamma^\alpha\,S\indices{^\beta_\alpha_\beta}\right)
+\frac{\rmi\ell}{6}\bsigma^{\alpha\beta}\bk_{\alpha\beta}\right]\psi=0.
\label{eq:gen-dirac-equation}
\end{align}
\section{Special cases: metric compatibility and zero torsion}
Based on the general Dirac equation~(\ref{eq:gen-dirac-equation}), we can derive special cases that apply for ``metric compatibility'' and beyond that, for vanishing torsion.
For the metric-compatible case we use the compatible spin connection, so that the tetrad postulate implies $\bgamma\indices{^\mu_;_\alpha}=0$.
It follows that $\bsigma\indices{^\mu^\nu_;_\alpha}=0$ and $(\sqrt{-g})\indices{_;_\alpha}=0$.
The generalized Dirac equation thus simplifies for metric compatibility, i.e., \mbox{$g_{\mu\nu;\xi}\equiv0$}, to
\begin{align}
\left[\rmi\bgamma^{\beta}+\frac{\rmi\ell}{3}\left(2\bsigma^{\xi\beta}S\indices{^\alpha_\xi_\alpha}-\bsigma^{\xi\alpha}S\indices{^\beta_\xi_\alpha}\right)\right]
\rmD_{\beta}\psi-\left(m+\rmi\bgamma^{\beta}S\indices{^\alpha_\beta_\alpha}
+\frac{\rmi\ell}{6}\bsigma^{\alpha\beta}\bk_{\alpha\beta}\right)\psi=0.
\label{eq:gen-dirac-equation1}
\end{align}
For metric compatibility, the curvature spinor-tensor $\bk_{\alpha\beta}$ is related to the Riemann-Cartan curvature tensor via (see~\ref{app0-gd}):
\begin{equation*}
\bk_{\alpha\beta}=-\iquarter\bsigma^{\xi\eta}R\indices{_{\xi\eta\alpha\beta}}
=\quarter\bgamma^{\xi}\bgamma^{\eta}R\indices{_{\xi\eta\alpha\beta}}\,.
\end{equation*}
Due to the skew-symmetry of the Riemann-Cartan curvature tensor in its last index pair, we have
\begin{equation}\label{eq:k-R-corr3}
\bsigma^{\alpha\beta}\bk_{\alpha\beta}=\iquarter\bgamma^{\alpha}\bgamma^{\beta}\bgamma^{\xi}\bgamma^{\eta}R\indices{_{\xi\eta\alpha\beta}}\,.
\end{equation}
With Eq.~(\ref{eq:k-R-corr3}), the generalized Dirac equation~(\ref{eq:gen-dirac-equation}) takes on the form:
\begin{align}
&\left[\rmi\bgamma^{\beta}+\frac{\rmi\ell}{3}\left(2\bsigma^{\xi\beta}S\indices{^\alpha_\xi_\alpha}
-\bsigma^{\xi\alpha}S\indices{^\beta_\xi_\alpha}\right)\right]\rmD_{\beta}\psi\nonumber\\
-&\left[m\,\bEins+\rmi\bgamma^{\beta}S\indices{^\alpha_\beta_\alpha}-\frac{\ell}{24}R\indices{_{\xi\eta\alpha\beta}}\,
\bgamma^{\alpha}\bgamma^{\beta}\bgamma^{\xi}\bgamma^{\eta}\right]\psi=0.
\label{eq:gen-dirac-equation2}
\end{align}
The equation contains explicit torsion-dependent derivative terms as well as torsion in the covariant derivative and a spin-curvature coupling.
The latter is matrix-valued for general Riemann--Cartan curvature; its reduction to a scalar mass coefficient uses the additional torsion-free symmetries below.

The Gasiorowicz and hence the Pauli-type coupling terms disappear setting $\ell=0$, whereby Eq.~(\ref{eq:gen-dirac-equation1}) reduces to
\begin{equation}\label{eq:gen-dirac-equation1a}
\rmi\bgamma^{\beta}\,\rmD_{\beta}\psi-\left(m\,\bEins+S\indices{^{\,\alpha}_{\beta\alpha}}\,\rmi\bgamma^{\beta}\right)\psi=0\,.
\end{equation}
This is the equation obtained from the symmetrized Dirac Lagrangian on a metric-compatible Riemann--Cartan background~\cite{frankel01,hehl76}.
The covariant derivative already contains contortion.
When it is separated into Levi-Civita and contortion contributions, its trace contribution cancels the explicit torsion-trace term in Eq.~(\ref{eq:gen-dirac-equation1a}).
The remaining minimal torsion coupling involves the totally antisymmetric part.
The displayed trace term must therefore not be interpreted as an independent additional interaction.

On the other hand, if torsion is neglected but $\ell \ne 0$, Eq.~(\ref{eq:gen-dirac-equation2}) simplifies to
\begin{equation}\label{eq:gen-dirac-equation3}
\rmi\bgamma^{\beta}\,\rmD_{\beta}\psi-
\left(m\,\bEins-\frac{\ell}{24}R\indices{_{\xi\eta\alpha\beta}}\,\bgamma^{\xi}\bgamma^{\eta}\bgamma^{\alpha}\bgamma^{\beta}\right)\psi=0\,.
\end{equation}
For $\ell=0$, this obviously reproduces again the conventional Dirac equation in curved spacetime.
As compared to the Dirac equation~(\ref{eq:dirac-equation}) in flat space, \emph{two} additional terms arise here:
(i) the well-known coupling of the spinor to the spinor connection $\spc_{\alpha}$~\cite{frankel01},
and (ii) a Pauli-type  term\footnote{We remark that this term also arises in the mechanical model
of relativistic spin~\cite{deriglasov17}.} that causes a direct coupling of the Riemann curvature tensor $R\indices{_{\xi\eta\alpha\beta}}$ with the spinor~$\psi$.
For the actual case of metric compatibility and zero torsion, this ``gravitational Pauli coupling term'' can simply be expressed in terms of the Ricci scalar
$R=R\indices{_{\xi\eta\alpha\beta}}\,g^{\xi\alpha}g^{\eta\beta}=R_{\eta\beta}g^{\eta\beta}$, as, by virtue of the symmetries of the Riemann tensor,
the following identity holds (see~\ref{app1-gd}):
\begin{equation}\label{eq:R-gamma-identity}
R\indices{_{\xi\eta\alpha\beta}}\,\bgamma^{\xi}\bgamma^{\eta}\bgamma^{\alpha}\bgamma^{\beta}=-2R\,\bEins\,.
\end{equation}
Equation~(\ref{eq:gen-dirac-equation3}) then finally simplifies to:
\begin{equation}\label{eq:gen-dirac-equation4a}
\rmi\bgamma^{\beta}\,\rmD_{\beta}\psi-\left(m+\frac{\ell}{12}R\right)\psi=0\,.
\end{equation}
Hence, in a torsion-free spacetime the ``gravitational Pauli coupling effect''
vanishes in regions with vanishing Ricci scalar \mbox{($R=0$)} of spacetime---in
particular for vacuum solutions of the Einstein equation with cosmological constant $\Lambda=0$.
Spin can source torsion in specified dynamical theories~\cite{vasak23}.
The torsion-free restriction used here is a background or test-field limit; compatibility with a fully coupled solution must be checked using the connection field equation of the selected gravitational action.
\section{Wavefront propagation of solutions of the generalized Dirac equation}\label{Sec:Propagation}
In order to investigate whether the generalized Dirac equation~(\ref{eq:gen-dirac-equation2})
maintains the causal properties of solutions of the conventional Dirac equation, we follow the analysis
of Velo and Zwanziger~\cite{PhysRev.186.1337,PhysRev.188.2218}.
The background is prescribed, smooth, real, and metric compatible; the coefficients are evaluated at a point.
This is an analysis of the highest spinor derivatives, not of the complete dynamical gravitational--fermionic system.
Let $n_\mu\ne0$ be a real characteristic covector and $u$ a nonzero leading discontinuity or geometric-optics spinor amplitude.
The terms of Eq.~(\ref{eq:gen-dirac-equation1}) that involve the highest order derivative are
\begin{equation*}
\left[\bgamma^{\beta}+\frac{\ell}{3}\left(2\bsigma^{\xi\beta}S\indices{^\alpha_\xi_\alpha}
-\bsigma^{\xi\alpha}S\indices{^\beta_\xi_\alpha}\right)\right]\rmi\rmD_{\beta}\psi=0.
\end{equation*}
Replacing $\rmi\rmD_\beta$ in the principal part by $n_\beta$ gives the amplitude equation~\cite[Chap.~VI]{courant62}
\begin{equation}
\left[\bgamma^{\beta}+\frac{\ell}{3}\left(2\bsigma^{\xi\beta}S\indices{^\alpha_\xi_\alpha}
-\bsigma^{\xi\alpha}S\indices{^\beta_\xi_\alpha}\right)\right]n_\beta\,u\equiv P(n)u=0,
\label{eq:gen-dirac-equation5}
\end{equation}
which defines the first-order principal symbol $P(n)$.
A nonzero amplitude exists if and only if $\det P(n)=0$.
We evaluate this condition through $\det(P(n)^2)=(\det P(n))^2$; equality of these zero sets does not make the squared matrix equation equivalent for every amplitude.
With the definition $V_\xi\coloneqq S\indices{^\alpha_\xi_\alpha}$
\begin{equation}\label{eq:gen-det}
\det\left(n^2\,\Eins+\frac{\rmi\ell}{3}\pmb{A}-\frac{\ell^2}{9}\pmb{B}\right)=0,
\end{equation}
where the spinor-matrix-valued spacetime scalars
\begin{align*}
\pmb{A}&=-\left(\bgamma^{\xi}\bgamma^{\alpha}\bgamma^{\eta}+\bgamma^{\eta}\bgamma^{\xi}\bgamma^{\alpha}\right)S\indices{^\beta_\xi_\alpha}n_\beta\,n_\eta\\
&=2\bgamma^{\xi}\left(2n^\alpha n^\beta\Eins-\bgamma^{\alpha}\bgamma^{\eta}n_\eta\,n^\beta\right)S\indices{_\beta_\xi_\alpha}.
\end{align*}
and
\begin{align*}
\pmb{B}&=4\left[{\left(V\cdot n\right)}^2-V^2 n^2\right]\Eins\\
&\quad+\left[\left(\bgamma^{\eta}\bgamma^{\tau}\bgamma^{\xi}\bgamma^{\alpha}-\bgamma^{\tau}\bgamma^{\eta}\bgamma^{\xi}\bgamma^{\alpha}
-\bgamma^{\xi}\bgamma^{\alpha}\bgamma^{\tau}\bgamma^{\eta}+\bgamma^{\xi}\bgamma^{\alpha}\bgamma^{\eta}\bgamma^{\tau}\right)V_\tau
+\bgamma^{\xi}\bgamma^{\alpha}\bgamma^{\tau}\bgamma^{\rho}S\indices{^\eta_\tau_\rho}\right] S\indices{^\beta_\xi_\alpha}n_\beta\,n_\eta,
\end{align*}
have been defined to simplify the expressions (see~\ref{app2-gd} for details).
The value of the determinant~(\ref{eq:gen-det}) depends on the particular $S\indices{^\beta_\xi_\alpha}$.
It can only be concluded that propagation might be non-causal depending on the specific torsion tensors.

We investigate in the following the particular cases of a totally antisymmetric torsion tensor and a pure vector torsion.
\subsection{Totally antisymmetric torsion tensor}
A totally antisymmetric torsion can be expressed in terms of the Levi-Civita tensor $\varepsilon_{\beta\xi\alpha\rho}$
by the axial vector $S^{\rho}$ via $S_{\beta\xi\alpha}\equiv\varepsilon_{\beta\xi\alpha\rho}\,S^\rho$.
Then the vector torsion vanishes: $V_\xi=0$.
When the length parameter is absorbed into $W^\rho \coloneqq \frac{2\ell}{3}S^{\rho}$,
the matrices $\pmb{A}$ and $\pmb{B}$ in Eq.~(\ref{eq:gen-det}) specialize to
\begin{align*}
\frac{\rmi\ell}{3}\pmb{A}&=-\rmi\varepsilon_{\beta\xi\alpha\rho}\,\bgamma^\xi\bgamma^\alpha\bgamma^\eta\,n^\beta n_\eta\,W^\rho\\
\frac{\ell^2}{9}\pmb{B}&=\quarter\varepsilon_{\beta\xi\alpha\rho}\,\varepsilon_{\eta\tau\lambda\sigma}\,
\bgamma^{\xi}\bgamma^{\alpha}\bgamma^{\tau}\bgamma^{\lambda}\,n^\beta n^\eta\,W^\rho W^\sigma\\
&=\left[W^2n^2-\left(W\cdot n\right)^2\right]\Eins.
\end{align*}
With the abbreviation
\begin{equation*}
C=n^2(1-W^2)+(W\cdot n)^2:=-\Delta\,\Eins,
\end{equation*}
the determinant condition~(\ref{eq:gen-det}) is then:
\begin{equation*}
0=\det\left(C\Eins+\frac{\rmi\ell}{3}\pmb{A}\right).
\end{equation*}
We define $\Delta=(W\cdot n)^2-W^2n^2$, so that $C=n^2+\Delta$, and write $\pmb L=\rmi\ell\pmb A/3$.
The explicit three-gamma expression implies $\tr\pmb L=0$.
The matrix square follows from the same Clifford and epsilon contractions:
\begin{equation*}
\frac{\ell^2}{9}\pmb A^2=-4n^2\Delta\Eins,
\qquad \pmb L^2=4n^2\Delta\Eins.
\end{equation*}
For clarity, setting
$\pmb E=\varepsilon_{\beta\xi\alpha\rho}\bgamma^\xi\bgamma^\alpha n^\beta W^\rho$
gives $\pmb E^2=-4\Delta\Eins$ and $\ell\pmb A/3=-\pmb E\bgamma^\eta n_\eta$.
The mixed ordering terms vanish because
\begin{equation*}
\bgamma^\eta n_\eta\pmb E-\pmb E\bgamma^\eta n_\eta
=2\varepsilon_{\beta\xi\alpha\rho}
(n^\xi\bgamma^\alpha-n^\alpha\bgamma^\xi)n^\beta W^\rho=0.
\end{equation*}
Thus the displayed square holds without omitting the term linear in $W^\rho$.

Put $\kappa=4n^2\Delta$. Tracelessness and $\pmb L^2=\kappa\Eins$ together imply
$\tr(\pmb L^2)=4\kappa$, $\tr(\pmb L^3)=0$, and $\tr(\pmb L^4)=4\kappa^2$.
For a traceless four-dimensional matrix, the determinant polynomial consequently gives
\begin{align*}
\det(C\Eins\pm\pmb L)
&=C^4-\frac12\tr(\pmb L^2)C^2
 \mathbin{\pm}\frac13\tr(\pmb L^3)C
 +\frac18[\tr(\pmb L^2)]^2-\frac14\tr(\pmb L^4)\\
&=(C^2-\kappa)^2.
\end{align*}
This proof includes $\kappa=0$ and establishes that the two determinant factors have the same zero set.
Their multiplication introduces neither an additional physical field equation nor additional characteristic covectors.
It yields
\begin{align*}
0&=\det\left(C\Eins+\frac{\rmi\ell}{3}\pmb A\right)
     \det\left(C\Eins-\frac{\rmi\ell}{3}\pmb A\right)\\
 &=\det\left(C^2\Eins+\frac{\ell^2}{9}\pmb A^2\right)
   =(C^2-4n^2\Delta)^4\\
 &=\left[(n^2+\Delta)^2-4n^2\Delta\right]^4
   =(n^2-\Delta)^8.
\end{align*}
The original squared-symbol determinant is $(n^2-\Delta)^4$, whereas $\det P(n)=(n^2-\Delta)^2$.
All three conditions have the same zeros, so the characteristic equation is
\begin{equation}\label{eq:char_det2}
(1+W^2)n^2-(W\cdot n)^2=0.
\end{equation}
Its polynomial form remains valid at $1+W^2=0$; only the equivalent quotient
$n^2=(W\cdot n)^2/(1+W^2)$ requires $1+W^2\ne0$.
Relative to the spacetime metric, causal characteristic hypersurfaces are timelike or null, with spacelike or null normals.
A timelike characteristic normal signals superluminal metric-relative wavefront propagation~\cite{Vijayalakshmi79,Fabbri:2018hhv}.
For $1+W^2>0$, Eq.~(\ref{eq:char_det2}) gives $n^2\geq0$; nonzero effective axial torsion admits timelike characteristic normals, although directions with $W\cdot n=0$ can remain null.
The zero-coupling case $W=0$ recovers the ordinary metric-null cone.
The distinction between the Lorentzian, degenerate, and ultrahyperbolic regimes is made explicitly below for both pure-torsion cases.
\subsection{Pure vector torsion}
A ``vector torsion'' $V_\xi$ is defined by a torsion tensor $S\indices{^\beta_\xi_\alpha}$ that satisfies the identity
\begin{equation*}
S\indices{^\beta_\xi_\alpha}\equiv \onethird\left( \delta^\beta_\alpha V_\xi-\delta^\beta_\xi V_\alpha\right).
\end{equation*}
The matrix $\pmb A$ then vanishes, and after absorbing the length parameter into $Z_\xi \coloneqq \frac{4\ell}{9}\,V_\xi$,  $B$ reduces to
\begin{equation*}
B=\frac{9}{\ell^2}\left({\left(n\cdot Z\right)}^2-n^2Z^2\right)\Eins.
\end{equation*}
The condition~(\ref{eq:gen-det}) now follows as
\begin{equation*}
\det P(n)=\left[\left(1+Z^2\right)n^2-{\left(Z\cdot n\right)}^2\right]^2=0,\qquad Z_\xi=\frac{4\ell}{9}V_\xi,
\end{equation*}
Thus
\begin{equation}\label{eq:vectortorsion}
(1+Z^2)n^2-(Z\cdot n)^2=0,
\end{equation}
with quotient $n^2=(Z\cdot n)^2/(1+Z^2)$ only for $1+Z^2\ne0$.
For nonzero timelike $Z$, the nonzero characteristic normals are timelike relative to the spacetime metric.
For spacelike $Z$, a negative denominator is not a restoration of ordinary causal propagation, as the quadratic-form classification shows.

\subsection{Characteristic type and scope of the propagation result}
For either pure-torsion case separately, write $U=W$ or $U=Z$ and define
\begin{equation*}
H_U^{\mu\nu}=(1+U^2)g^{\mu\nu}-U^\mu U^\nu,
\qquad Q_U(n)=H_U^{\mu\nu}n_\mu n_\nu.
\end{equation*}
The first-order determinant is $\det P(n)=Q_U(n)^2$.
These two pure-torsion results do not constitute a formula for an arbitrary superposition of torsion components.

In a local orthonormal frame, timelike $U^\mu=(u,0,0,0)$ gives
\begin{equation*}
Q_U(n)=\omega^2-(1+u^2)|\boldsymbol k|^2,
\qquad n_\mu=(\omega,\boldsymbol k).
\end{equation*}
For $u\ne0$, the front speed is $\sqrt{1+u^2}>1$.
For spacelike $U^\mu=(0,u,0,0)$,
\begin{equation*}
H_U^{\mu\nu}=\operatorname{diag}
\bigl(1-u^2,-1,-(1-u^2),-(1-u^2)\bigr).
\end{equation*}
When $|u|<1$, this form is Lorentzian and propagation parallel to $U$ has speed $1/\sqrt{1-u^2}>1$.
At $|u|=1$ the form degenerates; when $|u|>1$ it has signature $(2,2)$ and is ultrahyperbolic.
For nonzero null $U$, $Q_U(n)=n^2-(U\cdot n)^2$ is nondegenerate and Lorentzian, and it also admits timelike metric normals with special null directions.
Consequently, $1+U^2>0$ is the Lorentzian regime, $1+U^2=0$ is degenerate, and $1+U^2<0$ is ultrahyperbolic.
In the Lorentzian regime, nonzero effective torsion admits superluminal metric-relative fronts; it does not make every direction superluminal.

These are pointwise cone and characteristic-type statements for the spinor equation on prescribed backgrounds.
A Lorentzian characteristic polynomial alone is not a proof of strong hyperbolicity or well-posedness of a general first-order system; the eigenspaces, admissible initial hypersurfaces, and evolution estimates must also be analyzed.
For the coupled gravitational--fermionic system, one must specify the gravitational action, field equations, gauge treatment, and constraints.
In particular, the spin-curvature coupling contains derivatives of the connection, so those principal blocks cannot be inferred from the spinor block alone.

The displayed determinant is exact for the displayed classical equation.
If this Lagrangian is interpreted only as a leading effective interaction, its arbitrarily-high-frequency front limit requires additional qualification.
For example, the timelike axial result is $\sqrt{1+w^2}=1+\tfrac12w^2+O(w^4)$: the cone deformation begins at second order in the effective torsion coupling.
Operators omitted at the next order can affect propagation at the same order; no prediction of an unspecified ultraviolet completion is implied.
\section{Discussion}\label{Sec:Discussion}
The length parameter $\ell$ labels the selected covariantly regular free representative and becomes an interaction coefficient after coupling.
This construction supplies a relation among classical operators within the chosen model; it does not determine the numerical value of $\ell$ or show that all fermion species share it.
Likewise, the use of the same classical parameter for electromagnetic and gravitational interactions does not demonstrate that this relation is protected under quantum corrections.

The distinguishing feature of the present construction is the connection between covariant Legendre regularity and the electromagnetic, torsion, and curvature operators generated by the chosen coupling prescription.
Pauli-term contributions to anomalous magnetic moments have been investigated previously, including in Ref.~\cite{Sastry:1999is}.
Constraints on dipole operators in extensions of the Standard Model provide a broader context~\cite{Crivellin_2018,Alvarez_2023}.
An extension to electric dipole moments or flavor-changing decays such as $\mu\to e\gamma$ would require appropriate CP-odd or off-diagonal flavor coefficient structures.
They do not follow automatically from a real single-species magnetic Pauli coefficient.

Connections between $\ell$ and other scales, or its replacement by a scalar or composite field, are further model-building possibilities~\cite{Novello:2024kfi,Novello:2012ra}.
If $\ell$ becomes spacetime dependent, integrations by parts and the field equations acquire terms involving its derivatives; the constant-coefficient derivations here must then be reconsidered.
A Higgs-like symmetry-breaking potential would be one possible choice of additional dynamics, not a mathematical necessity of the substitution.
No dynamical mechanism fixing $\ell$ is supplied in the present work.

For a massless Dirac field, nonzero scalar curvature supplies a local scalar coefficient through $m_{\rm eff}=\ell R/12$ in the torsion-free equation.
A fixed coupling in the interacting Pauli and curvature--mass operators can break chiral symmetry explicitly; the dimensionful coupling also affects scale and conformal symmetry.
These are distinct symmetries, and no spontaneous symmetry-breaking mechanism is established here.
In free flat spacetime, the derivative-quadratic term remains a divergence and the usual massless bulk Dirac equation is unchanged.
The curvature--mass relation may be compared with Machian ideas about inertia and its gravitational environment~\cite{Mach1921,Sciama:1953zz,Novello:2011zza}, but it does not determine a particle's mass from the surrounding matter distribution.
Any torsional contribution to the gravitational environment depends on an additional gravitational model~\cite{Vasak:2023ncz}.

The effect on wave packets must be separated from the characteristic result.
For constant or slowly varying $m_{\rm eff}$, a local dispersion approximation gives
\begin{equation*}
E^2=|\boldsymbol p|^2+m_{\rm eff}^2,
\qquad |\boldsymbol v_{\rm group}|=|\boldsymbol p|/E.
\end{equation*}
The group speed is below unity for nonzero effective mass at finite momentum.
In contrast, the torsion-free principal symbol is $P(n)=\bgamma^\mu n_\mu$ and $\det P(n)=(n^2)^2$, independently of $m_{\rm eff}$; its wavefronts remain metric-null.
Returning to $R=0$ does not make a field with nonzero bare mass massless, and a suitably signed curvature contribution can in principle cancel a bare mass locally.

For the curvature convention of~\ref{app0-gd}, a de Sitter background is written as $R=-4\Lambda$ with $\Lambda>0$.
Equivalently, the vacuum Einstein-equation convention used for this benchmark is
$R_{\mu\nu}-\tfrac12 Rg_{\mu\nu}-\Lambda g_{\mu\nu}=0$.
This is a specified vacuum benchmark, not a field equation derived from the unspecified $\tilde{\LCd}_{\rm Gr}$.
Equation~(\ref{eq:gen-dirac-equation4a}) then reads
\begin{equation}\label{eq:gen-dirac-equation4}
\rmi\bgamma^\alpha
\left(\pfrac{\psi}{x^\alpha}+\spc_\alpha\psi\right)
-\left(m-\frac{\ell\Lambda}{3}\right)\psi=0.
\end{equation}
Once the action and charge conventions are fixed, the relative signs of the couplings have physical content; renaming the parameter does not remove this dependence.
Models in which an effective dark-energy term has torsional dynamics require their own coupled analysis~\cite{Vasak:2022gps,vasak20}.

For an explicit observational benchmark, Planck base-$\Lambda$CDM parameters give
$\Lambda\simeq1.1\times10^{-52}\,\mathrm{m}^{-2}=1.1\times10^{-82}\,\mathrm{fm}^{-2}$~\cite{Planck2018}.
Using the illustrative microscopic value $|\ell|=10^{-8}\,\mathrm{fm}$ and CODATA units~\cite{CODATA2022} gives
\begin{equation*}
|\Delta m|c^2=\frac{|\ell\Lambda|}{3}\hbar c
\simeq7\times10^{-83}\,\mathrm{eV}.
\end{equation*}
This de Sitter contribution is formally nonzero but negligible on particle-physics scales; it does not provide a quantitative explanation of neutrino masses.
A neutrino application requires suitable fermionic degrees of freedom and an explicit flavor structure.
Three coefficients diagonal in the weak-interaction flavor basis do not by themselves generate mixing.
A possible extension could introduce a flavor matrix $L_f$, with
$M_{\rm eff}=M+RL_f/12$, but its Hermitian action and its relation to the weak-interaction flavor basis must be specified.
Charged-lepton bounds cannot be transferred to neutrino coefficients without an explicit universality assumption.

Curvature-dependent fermion masses have been used to study inflation in specified cosmological models~\cite{Benisty:2019jqz}.
This motivates further investigation of high-curvature matter--gravity configurations, not a general conclusion that the present coupling prevents singularities.
The torsion-free interaction is equivalent, up to a divergence, to an action contribution
\begin{equation*}
-\int d^4x\,\sqrt{-g}\,\frac{\ell R}{12}\bar\psi\psi,
\end{equation*}
whose variation also affects the gravitational equations.
A prescribed local mass coefficient alone establishes neither repulsive collapse dynamics nor a curvature bound, bounce, or geodesic completeness.
Moreover, large tidal curvature or another divergent curvature invariant need not imply a large Ricci scalar, to which this particular torsion-free operator couples.
Claims about cosmological or black-hole singularity avoidance require a specified gravitational action, coupled equations, suitable solutions, and stability analysis.
\section{Summary and conclusions}\label{Sec:Conclusions}
Minimal coupling of the selected covariantly regular Dirac representative produces electromagnetic and gravitational Pauli-type interactions.
Its free derivative-quadratic term is a divergence, whereas the coupled contribution is physically nontrivial.
For electromagnetism, the action is equivalent up to a boundary term to the conventional Dirac--Maxwell action with Pauli coefficient $-q\ell/6$.
Matching the operator gives $|a_\ell|=2m|\ell|/3$; numerical constraints depend on the adopted magnetic-moment inputs and their uncertainty treatment.
The historical central-value scales in Section~\ref{sec:bound} are not confidence-level bounds, and the stated updated muon example is conditional on its specified hadronic and statistical inputs.

The preference for the regular representative is a structural assumption motivated by covariant canonical transformation theory~\cite{struckmeier08,struckmeier13,koenigstein16}.
Functional renormalization-group evidence motivates investigation of Pauli interactions~\cite{Gies:2020xuh,GiesTam2024}, but the present construction does not prove a necessary fundamental Pauli coefficient or ultraviolet completeness of QED.

In metric-compatible geometry, the additional terms include torsion-dependent derivative couplings and a spin-curvature coupling.
The latter becomes the scalar contribution $\ell R/12$ when torsion vanishes.
The background characteristic analysis gives
$\det P(n)=[(1+U^2)n^2-(U\cdot n)^2]^2$ for each of the two pure-torsion cases separately, with $U=W$ or $U=Z$.
It identifies superluminal metric-relative directions in the nonzero-effective-torsion Lorentzian regime and a change to degenerate and ultrahyperbolic characteristic forms at sufficiently large spacelike effective torsion.
This is not a complete well-posedness or characteristic analysis of a coupled dynamical gravitational--fermionic system.

The observed de Sitter vacuum contribution is negligible for the illustrative microscopic coupling.
The construction provides a basis for studying a specified coupled theory, including within covariant canonical gauge gravity~\cite{vasak23,Vasak:2018gqn,struckmeier21a} or related first-order approaches~\cite{Sciama:1953zz,kibble61,hehl76,hehl14}.
It does not by itself establish neutrino mixing, a universal curvature bound, or singularity avoidance.
\ack
The authors are indebted to the ``Walter Greiner-Gesellschaft e.V.'' in Frankfurt for support.
DV thanks especially the Fueck Stiftung for continuous support.
We thank F.W.~Hehl (Universit\"at K\"oln) for helpful comments
and appreciate the inspiring comments and constructive suggestions  of the unknown referees.
\appendix
\section{\texorpdfstring{Correlation of the curvature spinor $\br_{\alpha\beta}$ to the Riemann--Cartan tensor $R\indices{^{\eta}_{\xi\alpha\beta}}$}{Correlation of spin curvature and the Riemann--Cartan tensor}\label{app0-gd}}
For the spinor gauge theory, it is necessary to work out the correlation
$\br_{\alpha\beta}\leftrightarrow R\indices{^{\eta}_{\xi\alpha\beta}}$.
The spin curvature represents the Lorentz curvature in the spinor representation.
Coordinate indices are Greek; the Latin $i,j$ used in the tetrad construction are tangent-space Lorentz indices, distinct from the suppressed spinor matrix indices.

With metric compatibility and the compatible spin connection specified in Section~7, the tetrad postulate gives $\bgamma_{\mu;\alpha}=0$, hence
\begin{equation}\label{eq:covdiff-gamma}
\pfrac{\bgamma_{\mu}}{x^{\alpha}}=\bgamma_{\mu}\spc_{\alpha}-\spc_{\alpha}\bgamma_{\mu}+\bgamma_{\xi}\Gamma\indices{^{\xi}_{\mu\alpha}}.
\end{equation}
In order to work out the correlation of the Riemann tensor and the curvature spinor, defined in
Eq.~(\ref{eq:curvature-spinor}), we calculate the derivative of Eq.~(\ref{eq:covdiff-gamma})
\begin{equation*}
\ppfrac{\bgamma_{\mu}}{x^{\alpha}}{x^{\beta}}
=\pfrac{\bgamma_{\mu}}{x^{\beta}}\spc_{\alpha}+\bgamma_{\mu}\pfrac{\spc_{\alpha}}{x^{\beta}}
-\pfrac{\spc_{\alpha}}{x^{\beta}}\bgamma_{\mu}-\spc_{\alpha}\pfrac{\bgamma_{\mu}}{x^{\beta}}
+\pfrac{\bgamma_{\xi}}{x^{\beta}}\Gamma\indices{^{\xi}_{\mu\alpha}}+\bgamma_{\xi}\pfrac{\Gamma\indices{^{\xi}_{\mu\alpha}}}{x^{\beta}}.
\end{equation*}
Swapping the sequence of differentiation, the difference is encountered as
\begin{align*}
0&=\bgamma_{\mu}\left(\pfrac{\spc_{\alpha}}{x^{\beta}}-\pfrac{\spc_{\beta}}{x^{\alpha}}\right)-
\left(\pfrac{\spc_{\alpha}}{x^{\beta}}-\pfrac{\spc_{\beta}}{x^{\alpha}}\right)\bgamma_{\mu}+
\bgamma_{\xi}\left(\pfrac{\Gamma\indices{^{\xi}_{\mu\alpha}}}{x^{\beta}}-\pfrac{\Gamma\indices{^{\xi}_{\mu\beta}}}{x^{\alpha}}\right)\\
&\quad+\pfrac{\bgamma_{\mu}}{x^{\beta}}\spc_{\alpha}-\pfrac{\bgamma_{\mu}}{x^{\alpha}}\spc_{\beta}
-\spc_{\alpha}\pfrac{\bgamma_{\mu}}{x^{\beta}}+\spc_{\beta}\pfrac{\bgamma_{\mu}}{x^{\alpha}}
+\pfrac{\bgamma_{\xi}}{x^{\beta}}\Gamma\indices{^{\xi}_{\mu\alpha}}-\pfrac{\bgamma_{\xi}}{x^{\alpha}}\Gamma\indices{^{\xi}_{\mu\beta}}.
\end{align*}
Inserting Eq.~(\ref{eq:covdiff-gamma}) for the first derivatives of $\bgamma_{\mu}$ yields
\begin{align*}
0&=\bgamma_{\mu}\left(\pfrac{\spc_{\alpha}}{x^{\beta}}-\pfrac{\spc_{\beta}}{x^{\alpha}}\right)-
\left(\pfrac{\spc_{\alpha}}{x^{\beta}}-\pfrac{\spc_{\beta}}{x^{\alpha}}\right)\bgamma_{\mu}+
\bgamma_{\xi}\left(\pfrac{\Gamma\indices{^{\xi}_{\mu\alpha}}}{x^{\beta}}-\pfrac{\Gamma\indices{^{\xi}_{\mu\beta}}}{x^{\alpha}}\right)\\
&\quad+\left(\bgamma_{\mu}\spc_{\beta}-\xcancel{\spc_{\beta}\bgamma_{\mu}}+\cancel{\bgamma_{\xi}\Gamma\indices{^{\xi}_{\mu\beta}}}\,\right)\spc_{\alpha}-
\left(\bgamma_{\mu}\spc_{\alpha}-\xcancel{\spc_{\alpha}\bgamma_{\mu}}+\bcancel{\bgamma_{\xi}\Gamma\indices{^{\xi}_{\mu\alpha}}}\,\right)\spc_{\beta}\\
&\quad-\spc_{\alpha}\left(\xcancel{\bgamma_{\mu}\spc_{\beta}}-\spc_{\beta}\bgamma_{\mu}+\cancel{\bgamma_{\xi}\Gamma\indices{^{\xi}_{\mu\beta}}}\,\right)
+\spc_{\beta}\left(\xcancel{\bgamma_{\mu}\spc_{\alpha}}-\spc_{\alpha}\bgamma_{\mu}+\bcancel{\bgamma_{\xi}\Gamma\indices{^{\xi}_{\mu\alpha}}}\,\right)\\
&\quad+\left(\bcancel{\bgamma_{\xi}\spc_{\beta}}-\bcancel{\spc_{\beta}\bgamma_{\xi}}+\bgamma_{\eta}\Gamma\indices{^{\eta}_{\xi\beta}}\right)\Gamma\indices{^{\xi}_{\mu\alpha}}-
\left(\cancel{\bgamma_{\xi}\spc_{\alpha}}-\cancel{\spc_{\alpha}\bgamma_{\xi}}+\bgamma_{\eta}\Gamma\indices{^{\eta}_{\xi\alpha}}\right)\Gamma\indices{^{\xi}_{\mu\beta}}.
\end{align*}
Rearranging the terms gives
\begin{align*}
0&=\bgamma_{\mu}\left(\pfrac{\spc_{\alpha}}{x^{\beta}}-\pfrac{\spc_{\beta}}{x^{\alpha}}+\spc_{\beta}\,\spc_{\alpha}-\spc_{\alpha}\,\spc_{\beta}\right)-
\left(\pfrac{\spc_{\alpha}}{x^{\beta}}-\pfrac{\spc_{\beta}}{x^{\alpha}}+\spc_{\beta}\,\spc_{\alpha}-\spc_{\alpha}\,\spc_{\beta}\right)\bgamma_{\mu}\\
&\quad+\bgamma_{\xi}\left(\pfrac{\Gamma\indices{^{\xi}_{\mu\alpha}}}{x^{\beta}}-\pfrac{\Gamma\indices{^{\xi}_{\mu\beta}}}{x^{\alpha}}+
\Gamma\indices{^{\xi}_{\eta\beta}}\Gamma\indices{^{\eta}_{\mu\alpha}}-
\Gamma\indices{^{\xi}_{\eta\alpha}}\Gamma\indices{^{\eta}_{\mu\beta}}\right),
\end{align*}
while all other terms cancel.
With the Riemann-Cartan tensor $R\indices{^{\xi}_{\mu\beta\alpha}}$, defined by the sum proportional to $\bgamma_{\xi}$,
and the curvature spinor~(\ref{eq:curvature-spinor}), the equation writes concisely
\begin{equation}\label{eq:k-R-corr0}
\bgamma_{\mu}\br_{\beta\alpha}-\br_{\beta\alpha}\bgamma_{\mu}+\bgamma_{\xi}R\indices{^\xi_{\mu\beta\alpha}}=0.
\end{equation}
Equation~(\ref{eq:k-R-corr0}) is identically satisfied by
\begin{equation*}
\br_{\beta\alpha}=-\iquarter\bsigma\indices{_\xi^\eta}R\indices{^{\xi}_{\eta\beta\alpha}}
\quad\Leftrightarrow\quad R\indices{^\xi_{\mu\beta\alpha}}=\ihalf\Tr\left\{\bsigma\indices{^\xi_\mu}\,\br_{\beta\alpha}\right\},
\end{equation*}
as can be seen by contracting Eq.~(\ref{eq:k-R-corr0}) with $\bgamma^\eta$ and taking the trace of the spinor indices:
\begin{align*}
0&=\Tr\left\{\bgamma_\mu\br_{\beta\alpha}\bgamma^\eta-\br_{\beta\alpha}\bgamma_\mu\bgamma^\eta\right\}
+\Tr\left\{\bgamma_\xi\bgamma^\eta\right\}R\indices{^\xi_{\mu\beta\alpha}}\\
&=\Tr\left\{\left(\bgamma^\eta\bgamma_\mu-\bgamma_\mu\bgamma^\eta\right)\br_{\beta\alpha}\right\}
+\Tr\left\{\bgamma_\xi\bgamma^\eta\right\}R\indices{^\xi_{\mu\beta\alpha}}\\
&=-2\rmi\Tr\left\{\bsigma\indices{^\eta_\mu}\br_{\beta\alpha}\right\}+4\delta_\xi^\eta\,R\indices{^\xi_{\mu\beta\alpha}}.
\end{align*}
The last line uses the Clifford trace identities
$\Tr(\bgamma_\xi\bgamma^\eta)=4\delta_\xi^\eta$ and the definition of $\bsigma^{\eta}{}_{\mu}$.
The relation for $\br_{\beta\alpha}$ assumes the Lorentz-algebra-valued spin connection of Eq.~(\ref{eq:spinorconnection}).
The equation involving its products with $\bgamma_\mu$ alone would allow an additional identity-matrix curvature term; that central freedom is excluded here by the pure Lorentz restriction.
A simultaneously included U$(1)$ connection would have to be separated explicitly.
\section{Full contraction of the Riemann tensor with Dirac matrices\label{app1-gd}}
From the definition of the Dirac algebra for a general contravariant metric $g^{\eta\alpha}(x)=g^{\alpha\eta}(x)$,
\begin{equation*}
\onehalf\left(\gamma\indices{^a_c^\eta}\,\gamma\indices{^c_b^\alpha}+\gamma\indices{^a_c^\alpha}\,\gamma\indices{^c_b^\eta}\right)=
g^{\eta\alpha}\,\delta_b^a\quad\Longleftrightarrow\quad
\onehalf\left(\bgamma^{\eta}\bgamma^{\alpha}+\bgamma^{\alpha}\bgamma^{\eta}\right)=g^{\eta\alpha}\,\bEins,
\end{equation*}
where the Latin spinor indices refer to the spinor space and $\bEins$ denotes the unit matrix in spinor space, one concludes
\begin{align*}
\bgamma^{\eta}\bgamma^{\alpha}\bgamma^{\beta}&=
\left(\bgamma^{\eta}\bgamma^{\alpha}+\bgamma^{\alpha}\bgamma^{\eta}\right)\bgamma^{\beta}-\bgamma^{\alpha}\bgamma^{\eta}\bgamma^{\beta}\\
&=2g^{\eta\alpha}\bgamma^{\beta}-\bgamma^{\alpha}\bgamma^{\eta}\bgamma^{\beta}\\
&=2g^{\eta\alpha}\bgamma^{\beta}-\bgamma^{\alpha}\left(\bgamma^{\eta}\bgamma^{\beta}+\bgamma^{\beta}\bgamma^{\eta}\right)+
\bgamma^{\alpha}\bgamma^{\beta}\bgamma^{\eta}\\
&=2g^{\eta\alpha}\bgamma^{\beta}-2g^{\beta\eta}\bgamma^{\alpha}+\left(\bgamma^{\alpha}\bgamma^{\beta}+
\bgamma^{\beta}\bgamma^{\alpha}\right)\bgamma^{\eta}-\bgamma^{\beta}\bgamma^{\alpha}\bgamma^{\eta}\\
&=2g^{\eta\alpha}\bgamma^{\beta}-2g^{\beta\eta}\bgamma^{\alpha}+2g^{\alpha\beta}\bgamma^{\eta}-\bgamma^{\beta}\bgamma^{\alpha}\bgamma^{\eta},
\end{align*}
hence
\begin{equation}\label{eq:three-gamma}
\bgamma^{\eta}\bgamma^{\alpha}\bgamma^{\beta}+\bgamma^{\beta}\bgamma^{\alpha}\bgamma^{\eta}
=2\left(g^{\eta\alpha}\bgamma^{\beta}-g^{\beta\eta}\bgamma^{\alpha}+g^{\alpha\beta}\bgamma^{\eta}\right).
\end{equation}
By virtue of the skew-symmetry of the Riemann tensor in both its first and last index pair
\begin{equation*}
R_{\xi\eta\alpha\beta}=-R_{\eta\xi\alpha\beta},\qquad R_{\xi\eta\alpha\beta}=-R_{\xi\eta\beta\alpha},
\end{equation*}
the contraction $R_{\xi\eta\alpha\beta}\,\bgamma^{\xi}\bgamma^{\eta}\bgamma^{\alpha}\bgamma^{\beta}$ yields
\begin{align*}
R_{\xi\eta\alpha\beta}\,\bgamma^{\xi}\bgamma^{\eta}\bgamma^{\alpha}\bgamma^{\beta}&=\onehalf R_{\xi\eta\alpha\beta}\left[
\bgamma^{\xi}\left(\bgamma^{\eta}\bgamma^{\alpha}\bgamma^{\beta}\right)+
\left(\bgamma^{\xi}\bgamma^{\eta}\bgamma^{\alpha}\right)\bgamma^{\beta}\right]\\
&=R_{\xi\eta\alpha\beta}\left[\bgamma^{\xi}\left(
g^{\eta\alpha}\bgamma^{\beta}-g^{\beta\eta}\bgamma^{\alpha}+\cancel{g^{\alpha\beta}\bgamma^{\eta}}-
\onehalf\bgamma^{\beta}\bgamma^{\alpha}\bgamma^{\eta}\right)\right.\\
&\qquad\qquad+\left.\left(\,\bcancel{g^{\xi\eta}\bgamma^{\alpha}}-g^{\alpha\xi}\bgamma^{\eta}+
g^{\eta\alpha}\bgamma^{\xi}-\onehalf\bgamma^{\alpha}\bgamma^{\eta}\bgamma^{\xi}\right)\bgamma^{\beta}\right]\\
&=R_{\xi\eta\alpha\beta}\left(2g^{\eta\alpha}\bgamma^{\xi}\bgamma^{\beta}-
g^{\beta\eta}\bgamma^{\xi}\bgamma^{\alpha}-g^{\alpha\xi}\bgamma^{\eta}\bgamma^{\beta}-
\onehalf\bgamma^{\xi}\bgamma^{\beta}\bgamma^{\alpha}\bgamma^{\eta}-
\onehalf\bgamma^{\alpha}\bgamma^{\eta}\bgamma^{\xi}\bgamma^{\beta}\right)\\
&=-4R_{\xi\eta\alpha\beta}\,g^{\xi\alpha}\,\bgamma^{\eta}\bgamma^{\beta}-\onehalf R_{\xi\eta\alpha\beta}\left(
\bgamma^{\xi}\bgamma^{\beta}\bgamma^{\alpha}\bgamma^{\eta}+
\bgamma^{\alpha}\bgamma^{\eta}\bgamma^{\xi}\bgamma^{\beta}\right).
\end{align*}
Making use of the algebraic Bianchi identity $R_{\xi[\eta\alpha\beta]}=0$ in the metric-compatible torsion-free case,
\begin{equation*}
R_{\xi\eta\alpha\beta}+R_{\xi\alpha\beta\eta}+R_{\xi\beta\eta\alpha}=0\qquad\Rightarrow\qquad R_{\eta\beta}=R_{\beta\eta},
\end{equation*}
hence
\begin{align*}
\left(R_{\xi\eta\alpha\beta}+R_{\xi\alpha\beta\eta}+R_{\xi\beta\eta\alpha}\right)\bgamma^{\xi}\bgamma^{\beta}\bgamma^{\alpha}\bgamma^{\eta}&=0\\
\quad\Leftrightarrow\quad R_{\xi\eta\alpha\beta}\left(\bgamma^{\xi}\bgamma^{\beta}\bgamma^{\alpha}\bgamma^{\eta}+
\bgamma^{\xi}\bgamma^{\alpha}\bgamma^{\eta}\bgamma^{\beta}+\bgamma^{\xi}\bgamma^{\eta}\bgamma^{\beta}\bgamma^{\alpha}\right)&=0,
\end{align*}
yields
\begin{align*}
-\onehalf R_{\xi\eta\alpha\beta}\left(\bgamma^{\xi}\bgamma^{\beta}\bgamma^{\alpha}\bgamma^{\eta}+
\bgamma^{\alpha}\bgamma^{\eta}\bgamma^{\xi}\bgamma^{\beta}\right)
=\onehalf R_{\xi\eta\alpha\beta}\left(\bgamma^{\xi}\bgamma^{\eta}\bgamma^{\beta}\bgamma^{\alpha}+
\bgamma^{\xi}\bgamma^{\alpha}\bgamma^{\eta}\bgamma^{\beta}-
\bgamma^{\alpha}\bgamma^{\eta}\bgamma^{\xi}\bgamma^{\beta}\right).
\end{align*}
The full contraction of the Riemann tensor with the Dirac matrices now simplifies to
\begin{align*}
R_{\xi\eta\alpha\beta}\,\bgamma^{\xi}\bgamma^{\eta}\bgamma^{\alpha}\bgamma^{\beta}&=
-2R_{\eta\beta}\!\left(\bgamma^{\eta}\bgamma^{\beta}+\bgamma^{\beta}\bgamma^{\eta}\right)+
\onehalf R_{\xi\eta\alpha\beta}\!\left(\bgamma^{\xi}\bgamma^{\eta}\bgamma^{\beta}\bgamma^{\alpha}+
\bgamma^{\xi}\bgamma^{\alpha}\bgamma^{\eta}\bgamma^{\beta}+\bgamma^{\alpha}\bgamma^{\xi}\bgamma^{\eta}\bgamma^{\beta}\right)\\
&=-4R_{\eta\beta}g^{\eta\beta}\bEins+
\onehalf R_{\xi\eta\alpha\beta}\left[-\bgamma^{\xi}\bgamma^{\eta}\bgamma^{\alpha}\bgamma^{\beta}+\left(
\bgamma^{\xi}\bgamma^{\alpha}+\bgamma^{\alpha}\bgamma^{\xi}\right)\bgamma^{\eta}\bgamma^{\beta}\right]\\
&=-4R\,\bEins-\onehalf R_{\xi\eta\alpha\beta}\,\bgamma^{\xi}\bgamma^{\eta}\bgamma^{\alpha}\bgamma^{\beta}+
R_{\xi\eta\alpha\beta}\,g^{\xi\alpha}\bgamma^{\eta}\bgamma^{\beta}\\
&=-3R\,\bEins-\onehalf R_{\xi\eta\alpha\beta}\,\bgamma^{\xi}\bgamma^{\eta}\bgamma^{\alpha}\bgamma^{\beta},
\end{align*}
and finally,
\begin{equation*}
R_{\xi\eta\alpha\beta}\,\bgamma^{\xi}\bgamma^{\eta}\bgamma^{\alpha}\bgamma^{\beta}=-2R\,\bEins
\quad\Leftrightarrow\quad R_{\xi\eta\alpha\beta}\,\gamma\indices{^a_c^\xi}\,
\gamma\indices{^c_d^\eta}\,\gamma\indices{^d_e^\alpha}\,\gamma\indices{^e_b^\beta}=
-2R_{\xi\eta\alpha\beta}\,g^{\xi\alpha}\,g^{\eta\beta}\,\delta_b^a.
\end{equation*}
\section{Characteristic determinant\label{app2-gd}}
The principal-amplitude equation~(\ref{eq:gen-dirac-equation5}) has a nonzero spinor amplitude for a nonzero covector $n_\mu$ if and only if its characteristic determinant vanishes:
\begin{equation}\label{eq:char_det1}
\det\left[\bgamma^{\beta}n_\beta+\frac{\ell}{3}\left(2\bsigma^{\xi\beta}S\indices{^\alpha_\xi_\alpha}-\bsigma^{\xi\alpha}S\indices{^\beta_\xi_\alpha}\right)n_\beta\right]=0,
\end{equation}
hence with the torsion vector, defined by $V_\xi\coloneqq S\indices{^\alpha_\xi_\alpha}$,
\begin{equation}
\det\left[\bgamma^{\beta}n_\beta+\frac{\rmi\ell}{3}\left(\bgamma^{\xi}\bgamma^{\beta}V_\xi
-\bgamma^{\beta}\bgamma^{\xi}V_\xi-\bgamma^{\xi}\bgamma^{\alpha}S\indices{^\beta_\xi_\alpha}\right)n_\beta\right]=0,
\label{eq:gen-char_det}
\end{equation}
For a square matrix $M$, $\det(M^2)=(\det M)^2$.
Thus the squared-symbol determinant below has the same characteristic covectors as the original first-order determinant, although the corresponding matrix equations need not admit the same arbitrary amplitudes:
\begin{align*}
\det\bigg\{&\left[\bgamma^{\beta}+\frac{\rmi\ell}{3}\left(\bgamma^{\xi}\bgamma^{\beta}V_\xi
-\bgamma^{\beta}\bgamma^{\xi}V_\xi-\bgamma^{\xi}\bgamma^{\alpha}S\indices{^\beta_\xi_\alpha}\right)\right]\\
\times&\left[\bgamma^{\eta}+\frac{\rmi\ell}{3}\Big(\bgamma^{\tau}\bgamma^{\eta}V_\tau-\bgamma^{\eta}\bgamma^{\tau}V_\tau
-\bgamma^{\tau}\bgamma^{\rho}S\indices{^\eta_\tau_\rho}\,\Big)\right]n_\beta\,n_\eta\bigg\}=0.
\end{align*}
Abbreviated with the spinor-matrix-valued spacetime scalars $\pmb A$ and $\pmb B$,
\begin{equation}
\det\left(n^2\,\Eins+\frac{\rmi\ell}{3}\pmb A-\frac{\ell^2}{9}\pmb B\right)=0,
\end{equation}
we see that for $\pmb A$ the terms proportional to $V$ cancel, leaving:
\begin{align*}
\pmb A&=-\left(\bgamma^{\xi}\bgamma^{\alpha}\bgamma^{\eta}+\bgamma^{\eta}\bgamma^{\xi}\bgamma^{\alpha}\right)S\indices{^\beta_\xi_\alpha}n_\beta\,n_\eta\\
&=2\bgamma^{\xi}\left(2n^\alpha n^\beta\Eins-\bgamma^{\alpha}\bgamma^{\eta}n_\eta\,n^\beta\right)S\indices{_\beta_\xi_\alpha}.
\end{align*}
The matrix $\pmb B$ is given in explicit form as
\begin{align*}
\pmb B&=\left(\bgamma^{\xi}\bgamma^{\beta}V_\xi-\bgamma^{\beta}\bgamma^{\xi}V_\xi-\bgamma^{\xi}\bgamma^{\alpha}S\indices{^\beta_\xi_\alpha}\right)
\Big(\bgamma^{\tau}\bgamma^{\eta}V_\tau-\bgamma^{\eta}\bgamma^{\tau}V_\tau-\bgamma^{\tau}\bgamma^{\rho}S\indices{^\eta_\tau_\rho}\,\Big)n_\beta\,n_\eta\\
&=\Big[\left(\bgamma^{\xi}\bgamma^{\beta}\bgamma^{\tau}\bgamma^{\eta}-\bgamma^{\xi}\bgamma^{\beta}\bgamma^{\eta}\bgamma^{\tau}
-\bgamma^{\beta}\bgamma^{\xi}\bgamma^{\tau}\bgamma^{\eta}+\bgamma^{\beta}\bgamma^{\xi}\bgamma^{\eta}\bgamma^{\tau}\right)V_\tau V_\xi\\
&\qquad-\left(\bgamma^{\tau}\bgamma^{\eta}\bgamma^{\xi}\bgamma^{\alpha}-\bgamma^{\eta}\bgamma^{\tau}\bgamma^{\xi}\bgamma^{\alpha}
+\bgamma^{\xi}\bgamma^{\alpha}\bgamma^{\tau}\bgamma^{\eta}-\bgamma^{\xi}\bgamma^{\alpha}\bgamma^{\eta}\bgamma^{\tau}\right)V_\tau S\indices{^\beta_\xi_\alpha}\\
&\qquad+\bgamma^{\xi}\bgamma^{\alpha}\bgamma^{\tau}\bgamma^{\rho}S\indices{^\beta_\xi_\alpha}S\indices{^\eta_\tau_\rho}\Big]n_\beta\,n_\eta.
\end{align*}
The sum proportional to $V_\tau V_\xi$ simplifies by virtue of the symmetry of the $\bgamma$ products in $\beta$ and $\eta$ and in $\tau$ and $\xi$:
\begin{align*}
\pmb B&=4\left({\left(V\cdot n\right)}^2-V^2 n^2\right)\Eins\\
&\quad+\left[\left(\bgamma^{\eta}\bgamma^{\tau}\bgamma^{\xi}\bgamma^{\alpha}-\bgamma^{\tau}\bgamma^{\eta}\bgamma^{\xi}\bgamma^{\alpha}
-\bgamma^{\xi}\bgamma^{\alpha}\bgamma^{\tau}\bgamma^{\eta}+\bgamma^{\xi}\bgamma^{\alpha}\bgamma^{\eta}\bgamma^{\tau}\right)V_\tau
+\bgamma^{\xi}\bgamma^{\alpha}\bgamma^{\tau}\bgamma^{\rho}S\indices{^\eta_\tau_\rho}\right] S\indices{^\beta_\xi_\alpha}n_\beta\,n_\eta.
\end{align*}

\end{document}